\documentclass[pra,aps,twocolumn,showpacs]{revtex4}

\usepackage{graphicx}
\usepackage{amsmath,amsfonts,amsbsy,amssymb,amsthm, xypic}
\usepackage{mathrsfs,bbm}
\newcommand{\dual}[1]{\langle #1\rvert}
\newcommand{\ket}[1]{\lvert#1\rangle}
\newcommand{\ip}[2]{\langle#1 \vert #2\rangle}
\newcommand{\op}[2]{\ket{#1}\dual{#2}}

\newcommand{\modsq}[1]{\lvert#1\rvert^2}

\newcommand{\cg}[3]{{^{#3}C^{#1}_{#2}}}
\newcommand{\ketbra}[2]{\op{#1}{#2}}
\newcommand{\braket}[2]{\ip{#1}{#2}}
\newcommand{\bra}[1]{\dual{#1}}
\newtheorem{id}{Identity}

\begin{document}
\title{Collective processes of an ensemble of spin-1/2 particles}
\author{Bradley A. Chase}
\email{bchase@unm.edu}
\author{JM Geremia}
\email{jgeremia@unm.edu}
\affiliation{Quantum Measurement \& Control Group, Department of Physics \& Astronomy, The University of New Mexico, Albuquerque, New Mexico 87131 USA}
\begin{abstract}
When the dynamics of a spin ensemble are expressible solely in terms of symmetric processes and collective spin operators, the \emph{symmetric collective states} of the ensemble are preserved.  These many-body states, which are invariant under particle relabeling, can be efficiently simulated since they span a subspace whose dimension is linear in the number of spins.  However, many open system dynamics break this symmetry, most notably when ensemble members undergo identical, but local, decoherence.  In this paper, we extend the definition of symmetric collective states of an ensemble of spin-1/2 particles in order to efficiently describe these more general \emph{collective processes}.  The corresponding \emph{collective states} span a subspace which grows quadratically with the number of spins.  We also derive explicit formulae for expressing arbitrary identical, local decoherence in terms of these states. 
\end{abstract}
\pacs{03.65.Fd,03.65.Yz,34.10.-x}
\maketitle
\section{Introduction}
The ability to model the open system dynamics of large spin ensembles is crucial to experiments that make use of many atoms, as is often the case in precision metrology \cite{Wineland1993,Geremia2005d,Romalis2003,Partner2007}, quantum information science \cite{Polzik2001,Kuzmich2003,Jessen2007} and quantum optical simulations of condensed matter phenomena \cite{Bloch2002,StamperKurn2006,Parkins2008}.  Unfortunately, the mathematical description of large atomic spin systems is complicated by the fact that the dimension of the Hilbert space $\mathscr{H}_N$ grows exponentially in the number of atoms $N$.  Realistic simulations of experiments quickly become intractable even for atom numbers smaller than $N \sim 10$.  Current experiments, however, often work with atom numbers of more than $N \sim 10^{10}$, meaning that direct simulation of these systems is well beyond feasible.   Moreover, simulations over a range $N \sim 1-10$ are far from adequate to discern even the qualitative behavior that would be expected in the $N \gg 1$ limit.  Fortunately, it is often the case that experiments involving large spin ensembles respect one or more dynamical symmetries that can be exploited to reduce the effective dimension of the ensemble's Hilbert space. One can then hope to achieve a sufficiently realistic model of experiments without an exponentially large description of the system.   

In particular, previous work has focused on the \emph{symmetric collective states} $\ket{\psi_S}$, which are invariant under the permutation of particle labels: $\hat{\Pi}_{ij}\ket{\psi_S} = \ket{\psi_S}$.  These states span the subspace $\mathscr{H}_S \subset\mathscr{H}_N$, which grows linearly with the number of particles, $\operatorname{dim}(\mathscr{H}_S) = Nj + 1$.  However, in order for $\mathscr{H}_S$ to be an invariant subspace, the dynamics of the system must be expressible solely in terms of \emph{symmetric processes}, which are particle permutation invariant, and \emph{collective operators}, which respect the irreducible representation structure of rotations on the spin ensemble.  Fortunately, even within this restrictive class, a wide variety of phenomenon may be observed, including spin-squeezing \cite{Kitagawa1993,Polzik1999} and zero-temperature phase transitions \cite{Parkins2008}.

In practice, symmetric atomic dynamics are achieved by ensuring that there is identical coupling between all the atoms in the ensemble and the electromagnetic fields (optical, magnetic, microwave, etc.) used to both drive and observe the system \cite{Stockton2003}.  This approximation can be quite good for all of the coherent dynamics, because with sufficient laboratory effort, electromagnetic intensities can be made homogeneous, ensuring that interactions do not distinguish between different atoms in the ensemble.  However, incoherent dynamics are often beyond the experimenter's control.  Although most types of decoherence are symmetric, they are not generally written using collective operators.  Instead they are expressed as identical Lindblad operators for each spin
\begin{equation} \label{Intro::SymmetricLindblad}
	\sum_{n=1}^{N} \biggl[\hat{a}^{(n)}\rho(\hat{a}^{(n)})^{\dagger} 
				- \frac{1}{2} (\hat{a}^{(n)})^{\dagger}\hat{a}^{(n)}\rho
				- \frac{1}{2} \rho(\hat{a}^{(n)})^{\dagger}\hat{a}^{(n)} \biggr] \ .
\end{equation}
  The fact that decoherence does not preserve $\mathscr{H}_S$ has been well appreciated and the standard practice in experiments that address the collective state of atomic ensembles has been either: (i) to model such experiments only in a very short-time limit where decoherence can be approximately ignored; or (ii) to use decoherence models that do respect the particle symmetry, but which are written using only collective operators.

In this paper, we focus on ensembles of spin-1/2 particles and generalize symmetric collective states to simply \emph{collective states}, $\ket{\psi_C}$, which span the space $\mathscr{H}_C$.  This subspace, for which $\mathscr{H}_S \subset \mathscr{H}_C \subset \mathscr{H}_N$, is invariant under \emph{collective processes}, which include symmetric decoherence of the form of Eq. \ref{Intro::SymmetricLindblad}.  We also give formulae for expressing arbitrary symmetric Lindblad operators in the collective state basis.  Since $\operatorname{dim}(\mathscr{H}_C) = O(N^2)$, this allows for efficient simulation of a broader class of collective spin dynamics and in particular, allows one to consider the effects of decoherence on previous simulations of symmetric collective spin states (see Ref. \cite{shortPaper}).  We note that dynamics symmetries for spin-1/2 particles have been studied in the context of decoherence-free quantum information processing \cite{Lidar1998, Bacon2001}.  Unlike our work, which uses symmetries to find a reduced description of a quantum system, these works seek to protect quantum information from decoherence by encoding within the degeneracies introduced by dynamical symmetries.

The remainder of the paper is organized as follows.  Section \ref{Section::GeneralStates} reviews the representation theory of the rotation group, which plays an important role in defining the symmetries related to $\mathscr{H}_S$ and $\mathscr{H}_C$.  Section \ref{Section::CollectiveStates} introduces collective states and Section \ref{Section::CollectiveProcesses} defines collective processes over these states. Section \ref{Section::MainResult} gives an identity for expressing arbitrary symmetric Lindblads, e.g. Eq. \ref{Intro::SymmetricLindblad}, over the collective states.  Section \ref{Section::Conclusion} concludes.
\section{General states of the ensemble} \label{Section::GeneralStates}
Consider an ensemble of $N$ spin-1/2 particles, with the $n^{th}$ spin characterized by its angular momentum $\hat{\mathbf{j}}^{(n)} = \{ \hat{j}_x^{(n)}, \hat{j}_y^{(n)}, \hat{j}_z^{(n)} \}$.  States of the spin ensemble are elements of the composite Hilbert space
\begin{equation}
	\mathscr{H}_N = \mathscr{H}^{(1)} \otimes \mathscr{H}^{(2)} \otimes \cdots \otimes \mathscr{H}^{(N)} 
\end{equation}
with $\operatorname{dim}(\mathscr{H}_N) = 2^N$.  Pure states of the ensemble, $\ket{\psi} \in \mathscr{H}_N$, are written as
\begin{equation}
	\ket{\psi} = \sum_{m_1,m_2,\ldots, m_N} c_{m_1,m_2,\ldots, m_N} \ket{m_1, m_2, \ldots, m_N}
\end{equation}
with $m_n = \pm \frac{1}{2}$ and where
\begin{equation}
	\ket{m_1,m_2,\ldots, m_N} = \ket{\frac{1}{2}, m_1}_1 \otimes \ket{\frac{1}{2}, m_2}_2 \otimes \cdots \otimes \ket{\frac{1}{2}, m_N}_N
\end{equation}
satisfies
\begin{equation} \label{Equation::ProductStateKet}
	\hat{j}_z^{(n)}\ket{m_1,m_2,\ldots,m_N} = \hbar m_n\ket{m_1,m_2,\ldots,m_N} .
\end{equation}
When studying the open-system dynamics of the spin ensemble, one must generally consider the density operator
\begin{multline} \label{Equation::ProductStateRho}
	\hat{\rho}  =  \sum_{\substack{m_1,m_2,\ldots, m_N\\ m'_1,m'_2,\ldots,m'_N}} 
		 \rho_{m_1,m_2, \cdots, m_N ;  m_1^\prime, m_2^\prime \cdots ,m_N^\prime}   \\
		  \times \ket{m_1,m_2,\ldots,m_N}
		 	\dual{m_1^\prime,m_2^\prime,\ldots,m_N^\prime}
\end{multline}
States expanded as in Eqs. \ref{Equation::ProductStateKet} and \ref{Equation::ProductStateRho} are said to be written in the \emph{product basis}.
\subsection{Representations of the Rotation Group}
For a single spin-1/2 particle, a spatial rotation through the Euler angles $R=(\alpha,\beta,\gamma)$ is described by the rotation operator
\begin{equation}
	\hat{R}(\alpha,\beta,\gamma) = 
	e^{ - i \alpha \hat{j}_\mathrm{z} }
		e^{ - i \beta \hat{j}_\mathrm{y} }
	e^{ - i \gamma \hat{j}_\mathrm{z} }
\end{equation}
The basis kets $\ket{\frac{1}{2},m}$ for this particle therefore transform under the rotation $R$ according to
\begin{equation}
	\hat{R} \ket{\frac{1}{2},m'} = \sum_m \mathscr{D}^{\frac{1}{2}}_{m',m} (R) \ket{\frac{1}{2},m}
\end{equation}
where the matrices $\mathscr{D}^{\frac{1}{2}}(R)$ have the elements
\begin{equation}
	\mathscr{D}^{\frac{1}{2}}_{m',m} = \dual{\frac{1}{2},m'}  \hat{R}(\alpha,\beta,\gamma) \ket{\frac{1}{2},m} .
\end{equation}
The rotation matrices $\mathscr{D}^{\frac{1}{2}}(R)$ form a $2-$dimensional representation of the rotation group.

For the ensemble of $N$ spin-1/2 particles, each component of the ket $\ket{\psi} = \ket{m_1,m_2,\ldots,m_N}$ transforms separately under a rotation so that an arbitrary state transforms as
\begin{equation}
	\ket{\psi'} =  [\mathscr{D}^{\frac{1}{2}}(R)]^{\otimes N} \ket{\psi}.
\end{equation}
The rotation matrices $\mathscr{D}(R) = [\mathscr{D}^{\frac{1}{2}}(R)]^{\otimes N}$ provide a reducible representation for the rotation group but can be decomposed into irreducible representations (irreps) as  
\begin{equation} \label{Equation::IrreducibleRotations}
	\mathscr{D}(R) = \bigoplus_{J=J_\text{min}}^{J_\text{max}}\bigoplus_{i=1}^{d^J_N}\mathscr{D}^{J,i}(R) \ .
\end{equation}
The quantum number $i(J) = 1, 2, \ldots, d^J_N$ is used to distinguish between the
\begin{equation}
	d^J_N =\frac{N!(2J+1)}{(\frac{N}{2}-J)!(\frac{N}{2}+J+1)!} \ ,\ \  0 \leq J \leq \frac{N}{2}\\
\end{equation}
degenerate irreps with total angular momentum $J$ \cite{Mihailov1977}.  That is to say, $d^J_N$ is the number of ways one can combine $N$ spin-1/2 particles to obtain total angular momentum $J$.  The matrix elements of a given irrep $\mathscr{D}^{J,i}(R)$
\begin{equation}
	\mathscr{D}^{J,i}_{M,M'} (R) = \dual{J,M,i}  \mathscr{D}^{\frac{1}{2}}(R)^{\otimes N} \ket{J,M',i} 
\end{equation}
are written in terms of the total angular momentum eigenstates
\begin{eqnarray}
	\hat{\mathbf{J}}^2 \ket{J,M,i} & = & J(J+1) \ket{J,M,i} \\
	\hat{J}_\mathrm{z} \ket{J,M,i} & = & M \ket{J,M,i} 
\end{eqnarray}
with $\hat{J}_z = \sum_{n = 1}^N \hat{j}^{(n)}_z$, $J_{\text{max}} = \frac{N}{2}$ and
	\begin{equation}
		J_{\text{min}} = \begin{cases}
						  \frac{1}{2} & \text{$N$ odd}\\
						  0 & \text{$N$ even} \ .					
					  	  \end{cases}
	\end{equation}
It is important to note that degenerate irreps have \emph{identical} matrix elements, i.e. 
\begin{multline}
	\dual{J,M,i}  \mathscr{D}^{\frac{1}{2}}(R)^{\otimes N} \ket{J,M',i} \\
	  = \dual{J,M,i'}  \mathscr{D}^{\frac{1}{2}}(R)^{\otimes N} \ket{J,M',i'}
\end{multline}
for all $i,i'$.

In this representation, pure states are written as
\begin{equation} \label{Equation::IrrepKet}
	\ket{\psi} = \sum_{J = J_\text{min}}^{J_\text{max}}\sum_{M = -J}^{J}\sum_{i=1}^{d^J_N} c_{J,M,i}\ket{J,M,i}
\end{equation}
and mixed states as
\begin{multline} \label{Equation::IrrepRho}
	\hat{\rho} = \sum_{J,J' = J_\text{min}}^{J_\text{max}}
	\sum_{M,M' = -J,J'}^{J,J'}
	\sum_{i,i'=1}^{d^J_N,d^{J'}_N} \\
			\rho_{J,M,i;J',M',i'} \ketbra{J,M,i}{J',M',i'}
\end{multline}
States written in the form of Eqs. \ref{Equation::IrrepKet} or \ref{Equation::IrrepRho} are said to be written in the \emph{irrep basis}.  We stress that both the product and irrep bases can describe any arbitrary state in $\mathscr{H}_N$.
\section{Collective States} \label{Section::CollectiveStates}
While the representations in Section \ref{Section::GeneralStates} allow us to express any state of the ensemble of spin-1/2 particles, the irrep basis suggests a scenario in which we could restrict attention to a much smaller subspace of $\mathscr{H}_N$.  In particular, the irrep structure of the rotation group, as expressed in Eq. \ref{Equation::IrreducibleRotations}, indicates that rotations on the ensemble do not mix irreps and that degenerate irreps transform identically under a rotation.

Following this line of reasoning, we introduce the \emph{collective states}, $\ket{\psi_C}$, which span the sub-Hilbert space $\mathscr{H}_C \subset \mathscr{H}_N$.  Collective states have the property that degenerate irreps are identical; for pure states, $c_{J,M,i} = c_{J,M,i'}$ for all $i$ and $i'$.  We note that the symmetric collective states mentioned in the introduction are the collective states with $c_{J,M,i} = 0 $ unless $J=\frac{N}{2}$ and thus correspond to the largest $J$ value irrep.  We also note that
\begin{equation}
	\operatorname{dim}{\mathscr{H}_C} = 
		\begin{cases}
			\frac{1}{4}(N+3)(N+1), &\text{ if $N$ odd}\\
			\frac{1}{4}(N+2)^2, &\text{ if $N$ even}		
		\end{cases} \ .
\end{equation}  

Physically, the collective states reflect an inability to address different degenerate irreps of the same total $J$.  This new symmetry allows us to effectively ignore the quantum number $i$ and write
\begin{align}
	\ket{\psi_C} &= \sum_{J = J_\text{min}}^{J_\text{max}}
					   \sum_{M = -J}^{J}
					   \sum_{i=1}^{d^J_N} c_{J,M,i} \ket{J,M,i}\nonumber\\
			&= \sum_{J = J_\text{min}}^{J_\text{max}}\sum_{M = -J}^{J}
					\sqrt{d^J_N} c_{J,M,N} \ket{J,M}
\end{align}
where we have defined effective basis kets
\begin{equation} \label{Equation::EffectiveKet}
	\sqrt{d^J_N}c_{J,M,N}\ket{J,M,N} := \sum_{i=1}^{d^J_N} c_{J,M,i} \ket{J,M,i}
\end{equation}
with effective amplitude $c_{J,M,N} = c_{J,M,i}$ for all $i$ (since the $c_{J,M,i}$ are equal for collective states).  

At first glance, the factor of $\sqrt{d^J_N}$ may seem bizarre.  However, its presence enables us to apply standard spin-$J$ operators to the effective kets without having to explicitly refer to their constituent degenerate irrep kets $\ket{J,M,i}$.  In other words, $\ket{J,M,N}$ actually represents $d^J_N$ degenerate kets, each with identical probability amplitude coefficients.  But since the matrix elements of degenerate irreps of a given $J$ are identical, we need not evaluate them individually.  

As an example, consider a rotation operator $\hat{R}$ which necessarily respects the irrep structure of the rotation group.  Calculating the expectation value of $\hat{R}$ by expanding the collective state $\ket{\psi_C}$ in the irrep basis, we have
\begin{align}
	&\bra{\psi_C} \hat{R} \ket{\psi_C}  \nonumber\\
	&=  \sum_{J,J'}\sum_{M,M'}\sum_{i,i'} 
					c^{*}_{J,M,i}c_{J',M',i'}\bra{J,M,i} \hat{R}\ket{J',M',i'} 
						\label{Equation::CollectiveKetExpecation::First}\\
						              & = \sum_{J}\sum_{M,M'}\sum_{i} c^{*}_{J,M,i}c_{J,M',i}
																\bra{J,M,i} \hat{R}\ket{J,M',i}
						\label{Equation::CollectiveKetExpecation::Second}\\
							& = \sum_{J}\sum_{M,M'} d^J_N c^{*}_{J,M,1} c_{J,M',1}\bra{J,M,1} \hat{R} \ket{J,M',1}
						\label{Equation::CollectiveKetExpecation::Third}
\end{align}
where in going from Eq. \ref{Equation::CollectiveKetExpecation::First} to \ref{Equation::CollectiveKetExpecation::Second}, we set $J=J'$ and $i=i'$ since rotation group elements do not mix irreps.  In reaching Eq. \ref{Equation::CollectiveKetExpecation::Third}, we have further used the collective state property that $c_{J,M,i} = c_{J,M',i'} \forall i,i'$ and the rotation irrep property that $\bra{J,M,i} \hat{R} \ket{J,M',i} = \bra{J,M,i^{\prime}} \hat{R} \ket{J,M',i^{\prime}} \forall i,i'$. We arbitrarily chose $i=1$ as the representative element.

Equivalently, if we define
\begin{equation}
	\bra{J,M,N} \hat{R} \ket{J,M',N} = \bra{J,M,1} \hat{R} \ket{J,M',1}
\end{equation}
and use the effective amplitude definition, we can evaluate the expectation using the effective basis kets $\ket{J,M,N}$ directly: 
\begin{align}
	& \bra{\psi_C} \hat{R} \ket{\psi_C} \nonumber\\
	& = \sum_{J,J'}\sum_{M,M'} d^J_N c^{*}_{J,M,N}c_{J',M',N} 
					\bra{J,M,N} \hat{R} \ket{J',M',N}\\
					& = \sum_{J}\sum_{M,M'} d^J_N  c^{*}_{J,M,N}c_{J,M',N} \bra{J,M,N} \hat{R} \ket{J,M,N} \ . \label{Equation::CollectiveKetExpectation:Proper}
\end{align}
We see then that the $\sqrt{d^J_N}$ factor allows us to calculate expectations directly in the collective notation.  

We can similarly define collective state density operators, $\hat{\rho}_C$, which have the properties that (i) there are no coherences between different irrep blocks and (ii) degenerate irrep blocks have identical density matrix elements.  The second assumption again means we can effectively drop the index $i$, since $\rho_{J,M,i;J,M',i} = \rho_{J,M,i';J,M',i'}$ for any $i$ and $i'$.  This allows us to write
\begin{multline}
	\hat{\rho}_C = \sum_{J = J_\text{min}}^{J_\text{max}}
					\sum_{M,M'= -J}^J \\
					\rho_{J,M,N;J,M',N} \ketbra{J,M,N}{J,M',N}
\end{multline}
where the generalization of Eq. \ref{Equation::EffectiveKet} for effective density matrix elements is given by
\begin{multline} \label{Equation::EffectiveDensityMatrixElement}
	d^J_N\rho_{J,M,N;J,M',N}\op{J,M,N}{J,M',N} := \\
		\sum_{i=1}^{d^J_n} \rho_{J,M,i;J,M',i} \op{J,M,i}{J,M',i} \ .
\end{multline}
Just as for the effective kets, the factor of $d^J_N$ ensures expectations are correctly calculated using the standard spin-$J$ operators.  The density matrix has $\frac{1}{2}(N+3)(N+2)(N+1)$ elements.

We stress that Eq. \ref{Equation::EffectiveDensityMatrixElement} is \emph{different} than naively taking the outer product of the effective kets defined in Eq. \ref{Equation::EffectiveKet}.  That approach would involve outer products of kets between different, although degenerate, irreps.  Such terms are strictly forbidden by the first property of collective state density operators.  Instead, one should consider both the effective kets and effective density operators as representing $d^J_N$ identical copies of a spin-$J$ particle.  Eqs. \ref{Equation::EffectiveKet} and \ref{Equation::EffectiveDensityMatrixElement} are then just rules for relating pure or mixed states of this effective spin-$J$ particle to the true states in the irrep basis.

\section{Collective Processes} \label{Section::CollectiveProcesses}
We are now interested in describing quantum processes, $\mathcal{L}$, which preserve collective states, $\hat{\rho}'_C = \mathcal{L}\hat{\rho}_C$.  Writing this explicitly, we must have
\begin{align}
	& \sum_{J_1}\sum_{M_1,M'_1} d^{J_1}_N 
				\rho'_{J_1,M_1,N;J_1,M'_1,N} \ketbra{J_1,M_1,N}{J_1,M'_1,N} \nonumber\\
	& =			\sum_{J_2}\sum_{M_2,M'_2} d^{J_2}_N 
					\rho_{J_2,M_2,N;J_2,M'_2,N} 
					\mathcal{L}	\ketbra{J_2,M_2,N}{J_2,M'_2,N} \label{Equation::CollectivePreservingDynamics} .
\end{align}
If we define the action of $\mathcal{L}$ on collective density matrix elements as
\begin{equation}
	f^{J,M,M',N} = \mathcal{L}	\ketbra{J,M,N}{J,M',N}
\end{equation}
we immediately see that this action must be expressible as
\begin{equation} \label{Equation::SymmetricProcess}
	f^{J,M,M',N} = \sum_{J_1}\sum_{M_1,M_1'} \lambda^{J,M,M',N}_{J_1,M_1,M_1'} \ketbra{J_1,M_1,N}{J_1,M_1',N}
\end{equation}
in order for the equality in Eq. \ref{Equation::CollectivePreservingDynamics} to be met.  Here $\lambda^{J,M,M',N}_{J_1,M_1,M_1'}$ is an arbitrary function of its indices.  Any process which preserves collective states by satisfying Eq. \ref{Equation::SymmetricProcess} is a \emph{collective process}.

Examples of collective processes are those involving $\emph{collective angular momentum operators}$ $\{\hat{J}_x,\hat{J}_y, \ldots\}$ and more generally, arbitrary \emph{collective operators} $\hat{C} = \sum_{n=1}^N \hat{c}^{(n)}$.  Since collective operators correspond to precisely the rotations considered when defining the irrep structure of the rotation group, they can all be written as 
\begin{equation} \label{Equation::CollectiveOperator}
	\hat{C} = \sum_{J}\sum_{M,M'} c_{J,M,M'}\ketbra{J,M,N}{J,M',N} \ , 
\end{equation}
which cannot couple effective matrix elements with different $J$.  

However, the collective operators define a more restrictive class than an arbitrary collective process, which \emph{can} couple different $J$ blocks, so long as it does not create coherences between them.  In fact, if all operators are collective, then the symmetric collective states ($\ket{\psi_S}$) span an invariant subspace of the map.  This holds  even when considering Lindblad operators that are written in terms of collective operators, such as \emph{collective spontaneous emmission}
\begin{equation}
	\mathcal{L}_{\Gamma_\text{col}}\hat{\rho} = \Gamma \biggl[\hat{J}_{-}\hat{\rho}\hat{J}_{+} -
				\frac{1}{2}\hat{J}_{+}\hat{J}_{-}\hat{\rho}
				-\frac{1}{2}\hat{\rho}\hat{J}_{+}\hat{J}_{-}\biggr]\ , 
\end{equation}
which describes a process in which all spins of the ensemble emit together. 

In the following section, we demonstrate that a process of the form
\begin{equation} \label{Equation::GeneralSymmetric}
	f^{J,M,M',N} = \sum_{n = 1}^{N} \hat{a}^{(n)} \ketbra{J,M,N}{J,M',N} (\hat{a}^{(n)})^{\dagger} \ ,
\end{equation}
which cannot be written solely in terms of collective operators, is nonetheless a collective process.  This result was inspired by considering  
\emph{symmetric spontaneous emmission}
\begin{equation}
	\mathcal{L}_{\Gamma_\text{sym}}\hat{\rho} = \Gamma \biggl[
				\sum_{n = 1}^{N} \hat{j}_{-}^{(n)}\hat{\rho}\hat{j}_{+}^{(n)}
				- \frac{1}{2}\hat{j}_{+}^{(n)}\hat{j}_{-}^{(n)}\hat{\rho}
				-\frac{1}{2}\hat{j}_{+}^{(n)}\hat{j}_{-}^{(n)}\biggr] \ ,
\end{equation}
which corresponds to the spins emitting independently via identical processes.  Accordingly, we call processes which are invariant under exchanging particle labels \emph{symmetric processes}.  
\begin{figure}[b]
	\begin{equation} 
		\xymatrix{ N = 1 & 2 & 3 & 4 \ldots\\
		& & &                        1\times 2\\
		& & 1 \times \frac{3}{2}\ar[ur]\ar[dr] & \\
		& 1 \times 1 \ar[ur]\ar[dr] & & 3\times 1\\
		1 \times \frac{1}{2}\ar[ur]\ar[dr] & & 2\times\frac{1}{2} \ar@{=>}[ur]\ar@{=>}[dr]& \\
		& 1 \times 0 \ar[ur] & & 2\times 0\\
		}\nonumber 
	\end{equation}
	\caption{Degeneracy structure from adding spin-1/2 particles, labeled as $d^J_N \times J$} \label{fig:MomentumTree} 
\end{figure}
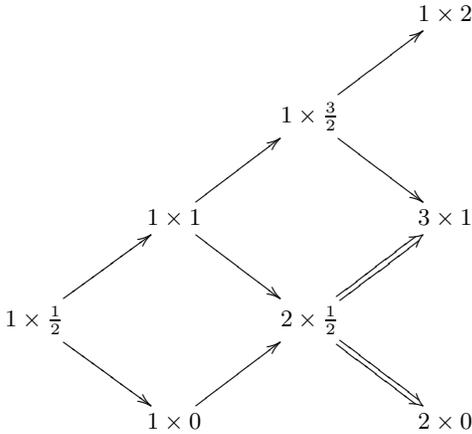

In order to appreciate how $\mathcal{L}_{\Gamma_\text{sym}}$ is related to Eq. \ref{Equation::GeneralSymmetric}, expand the single spin operator $\hat{a}$ in the spherical basis
\begin{equation}
	\hat{a} = a_I\hat{I} + \sum_{q} a_q\hat{j}_q = a_{I}\hat{I} + a_{+}\hat{j}_{+} + a_{-}\hat{j}_{-} + a_{z}\hat{j}_{z} 
\end{equation}
with the convention $\hbar = 1$, 
$\hat{j}_{+} = \bigl(\begin{smallmatrix}
		0 & 1\\
		0 & 0
	   \end{smallmatrix}\bigr)$,
$\hat{j}_{-} = \bigl(\begin{smallmatrix}
		0 & 0\\
		1 & 0
	   \end{smallmatrix}\bigr)$	and
$\hat{j}_{z} = \bigl(\begin{smallmatrix}
		1 & 0\\
		0 & -1
	   \end{smallmatrix}\bigr)$.  We can write an arbitrary symmetric Lindblad for $\hat{a}$, of which $\mathcal{L}_{\Gamma_\text{sym}}$ is a specific example, as
\begin{equation} \label{Equation::SymmetricLindblad}
	\begin{split}
		\mathcal{L}_{\hat{a}}\hat{\rho} &= \sum_{n=1}^{N}\biggr[ \hat{a}^{(n)}\hat{\rho}(\hat{a}^{(n)})^{\dagger} - 
					\frac{1}{2} (\hat{a}^{(n)})^{\dagger}\hat{a}^{(n)} \hat{\rho} -
					\frac{1}{2} \hat{\rho} (\hat{a}^{(n)})^{\dagger}\hat{a}^{(n)} \biggr]\\
					& = \sum_{n=1}^N \biggl[\hat{a}^{(n)}\hat{\rho}(\hat{a}^{(n)})^{\dagger}\biggr] -
							  \frac{1}{2} \hat{A}_N \hat{\rho}  - \frac{1}{2} \hat{\rho} \hat{A}_N
	\end{split}
\end{equation}
with the collective operator $\hat{A}_N$ given by
\begin{equation}
	\begin{split}
		\hat{A}_N &= \sum_{n=1}^N (\hat{a}^{(n)})^{\dagger}\hat{a}^{(n)} \\
		          &= \bigl(\frac{1}{2} \modsq{a_{-}} + 
						   \frac{1}{2} \modsq{a_{+}} +
						   \modsq{a_{I}}+ \modsq{a_z}\bigr)N\hat{I}\\
		          &+ \bigl( a_{-}^*a_I - a_{-}^{*}a_z + a_I^{*}a_{+} + a_z^{*}a_{+}\bigr)\hat{J}_{+}\\						
		 		  &+ \bigl( a_I^{*}a_{-} + a_{+}^{*}a_I + a_{+}^{*}a_Z - a_{z}^{*}a_{-} \bigr)\hat{J}_{-} \\
		          &+ \bigl( \frac{1}{2}\modsq{a_{-}} 
							- \frac{1}{2}\modsq{a_{+}} 
							+ a_I^{*}a_z + a_z^{*}a_I\bigr)\hat{J}_{z} \ .
	\end{split}
\end{equation}  
In this form, it is clear that only the first term of the symmetric Lindbladian is not written using collective operators.  In fact, if we again expand $\hat{a}^{(n)}$ in the spherical basis, we observe that the only terms which involve non-collective operators are those which do not involve the identity operator, 
\begin{equation} \label{Equation::NonCollectiveLindbladTerms}
	\begin{split}
		\sum_{n=1}^N \biggl[\hat{a}^{(n)}\hat{\rho}(\hat{a}^{(n)})^{\dagger}\biggr] &= \modsq{a_I}N\hat{\rho} + 
		\sum_{q}\bigl(a_qa_I^{*} \hat{J}_q\hat{\rho} + a_Ia_q^{*}\hat{\rho} \hat{J}_q^{\dagger}\bigr) \\
		& + \sum_{n=1}^N\biggl[\sum_{q,r}a_qa_r^{*} \hat{j}^{(n)}_q \hat{\rho} (\hat{j}^{(n)}_q)^{\dagger} \biggr]
	\end{split}
\end{equation}
with $\hat{J}_q = \sum_{n=1}^N \hat{j}^{(n)}_q$ a collective spin operator.  In the following section, we demonstrate that these processes preserve the collective states and we give an identity for evaluating such processes for arbitrary spherical basis elements.

\section{Identity} \label{Section::MainResult}
\begin{id} \label{Id::MainResult}
	Given a collective density matrix element for $N$ spin-1/2 particles, $\ketbra{J,M,N}{J,M',N}$, we have
	\begin{align}
		&f^{J,M,M',N}\\
		=&\sum_{n=1}^N \hat{j}_{q}^{(n)}\ketbra{J,M,N}{J,M',N}(\hat{j}_{r}^{(n)})^{\dagger}
				\label{Equation::BasicSum}\\
		=& \frac{1}{2J}\biggl[1+\frac{\alpha^{J+1}_N}{d^J_N}\frac{2J+1}{J+1}\biggr]
		A_q^{J,M}\ketbra{J,M_q,N}{J,M'_r,N} A_r^{J,M'} \nonumber \\
		+&\frac{\alpha^J_N}{d^J_N2J}B_q^{J,M}\ketbra{J-1,M_q,N}{J-1,M'_r,N} B_r^{J,M'} \nonumber\\
		+&\frac{\alpha^{J+1}_N}{d^{J}_N2(J+1)}D_q^{J,M}
		  \ketbra{J+1,M_q,N}{J+1,M'_r,N}D_r^{J,M'} \label{Equation::generalFormula}
	\end{align}
	where $q,r \in \{+,-,z\}$, $M_+ = M + 1$, $M_- = M - 1$ and $M_z = M$,
	\begin{equation}
		\alpha^J_N = \sum_{J'=J}^{\frac{N}{2}}d^{J'}_N = \frac{N!}{\left(\frac{N}{2}-J\right)! \left(\frac{N}{2}+J\right)!}
	\end{equation}

	and
	\begin{subequations}
		\begin{align}
			A_+^{J,M} &= \sqrt{(J-M)(J+M+1)}\\
			A_-^{J,M} &= \sqrt{(J+M)(J-M+1)}\\
			A_z^{J,M} &= M
		\end{align}
	\end{subequations}
	
	and
	\begin{subequations}
		\begin{align}
			B_+^{J,M} &=\sqrt{(J-M)(J-M-1)}\\
			B_-^{J,M} &=-\sqrt{(J+M)(J+M-1)} \\	
			B_z^{J,M} &= \sqrt{(J+M)(J-M)} 
		\end{align}
	\end{subequations}
	
	and lastly
	\begin{subequations}
		\begin{align}
			D_+^{J,M} &= -\sqrt{(J+M+1)(J+M+2)} \\
			D_-^{J,M} &=\sqrt{(J-M+1)(J-M+2)} \\	
			D_z^{J,M} &= \sqrt{(J+M+1)(J-M+1)} \ .
		\end{align}
	\end{subequations}
	
\end{id}
\noindent Note that $\alpha^J_N$ and $d^J_N$ are zero if $J$ is negative or $J = N/2$, ensuring that only valid density matrix elements are involved. 

In the following subsections, we prove Identity \ref{Id::MainResult} inductively. The motivation for the inductive proof comes from the simple recursive structure of adding spin-1/2 particles.  As seen in Fig. \ref{fig:MomentumTree}, the $d^J_N$ irreps which correspond to a total spin $J$ particle composed of $N$ spin-1/2 particles can be split into two groups, depending on how angular momentum was added to reach them.  By expressing the $N$ particle states in terms of bipartite states of a single spin-1/2 particle and a spin-$(N-1)$ particle, we can then evaluate the dynamics independently on either half by assuming Identity \ref{Id::MainResult} holds.  Returning the resulting state to the $N$ particle basis should then confirm the Identity.  By inspection, the base case of $N=1$ holds, as the $A_q^{J,M}$ terms reduce to the single spin-1/2 matrix elements.  We now proceed to the inductive case.
\subsection{Recursive state structure}
In order to apply the inductive hypothesis, we need to express an $N$ particle state in terms of $N-1$ particle states.  This recursive structure is best seen by examining Fig. \ref{fig:MomentumTree}, which illustrates the branching structure for adding spin-1/2 particles.  For example, the three-fold degenerate $N=4$ spin-1 irreps arise from two different spin additions---adding a single spin-1/2 particle to the non-degenerate $J=\frac{3}{2}$,$N=3$ irrep and adding to the 2-fold degenerate $J=\frac{1}{2}$, $N=3$ irreps. Since we are always adding a spin-1/2 particle, the tree is at most binary. This allows us to recursively decompose the degenerate irreps for a given $J$ in terms of adding a single spin-1/2 particle to the two related $N-1$ degenerate irreps.  

Recall that for the collective states, we define effective density matrix elements which group degenerate irreps (Eq. \ref{Equation::EffectiveDensityMatrixElement}).  In order to make the relationship between states of different $N$ more obvious, we rewrite the state $\ket{J,M,i}$ as $\ket{J,M,N,i}$.  The $N$ and $i$ indices indicate the state is from $i$-th degenerate total spin-$J$ irrep that comes from adding $N$ spin-1/2 particles. So that we can leverage the binary branching structure seen in Fig. \ref{fig:MomentumTree}, we need to then relate the $N$ particle irrep states to the $N-1$ particle irrep states. Accordingly, we define $\ket{J,M;\frac{1}{2},J\pm\frac{1}{2},N-1,i_1}$, where the last four entries indicate that the overall $N$ spin state can be viewed as combining a single spin-1/2 particle with a spin $J\pm \frac{1}{2}$ particle.  The spin $J\pm \frac{1}{2}$ particle is from the $i_1$-st such irrep for $N-1$ spin-1/2 particles.  With these definitions, we can now relate the $N$ particle states to the $N-1$ particle states by explicitly tensoring out a single spin-1/2 particle:
\begin{widetext}
	\begin{align} 
			 & \ketbra{J,M,N}{J,M',N} \\ 
			 =& \frac{1}{d^J_N}\sum_{i=1}^{d^J_N}\ketbra{J,M,N,i}{J,M',N,i}\\
			 =&\frac{1}{d^J_N}\sum_{i_1=1}^{d^{J+\frac{1}{2}}_{N-1}}
									\ketbra{J,M;\frac{1}{2},J+\frac{1}{2},N-1,i_1}
										   {J,M';\frac{1}{2},J+\frac{1}{2}, N-1,i_1} \nonumber\\
			+ & \frac{1}{d^J_N}\sum_{i_2=1}^{d^{J-\frac{1}{2}}_{N-1}}
			\ketbra{J,M;\frac{1}{2},J-\frac{1}{2},N-1,i_2}{J,M';\frac{1}{2},J-\frac{1}{2},N-1,i_2}\displaybreak[1]\\
			= & \frac{d^{J+\frac{1}{2}}_{N-1}}{d^J_N}
			 \sum_{m_1} \biggl[ \cg{\frac{1}{2},m_1}{J+\frac{1}{2},M-m_1}{J,M}
			\ket{\frac{1}{2},m_1}
			\ket{J+\frac{1}{2},M-m_1,N-1} \biggr]
		     \sum_{m_1'} \biggl[
		 		 \bra{\frac{1}{2},m_1'}
				\bra{J+\frac{1}{2},M'-m_1',N-1}
			 \cg{\frac{1}{2},m_1'}{J+\frac{1}{2},M'-m_1'}{J,M'} \biggr] \nonumber \\ 
			+ & \frac{d^{J-\frac{1}{2}}_{n-1}}{d^J_N}
			    \sum_{m_2} \biggl[ \cg{\frac{1}{2},m_2}{J-\frac{1}{2},M-m_2}{J,M}
		    \ket{\frac{1}{2},m_2}
		    \ket{J-\frac{1}{2},M-m_2,N-1} \biggr]
				\sum_{m_2'} \biggl[
		        \bra{\frac{1}{2},m_2'}
		        \bra{J-\frac{1}{2},M'-m_2',N-1}
		     \cg{\frac{1}{2},m_2'}{J-\frac{1}{2},M'-m_2'}{J,M'} \biggr] \label{eq:recursiveStateDefinition}
	\end{align}
\end{widetext}
with Clebsch-Gordan coefficients $\cg{j_1,m_1}{j_2,m_2}{J,M} = \braket{J,M; j_1, j_2}{j_1,m_1;j_2,m_2}$ and the $m_i,m_i'$ sums over single spin projection values $\pm \frac{1}{2}$.  In reaching Eq. \ref{eq:recursiveStateDefinition}, we made use of the definition of the effective density matrix element for $N-1$ spins given in Eq. \ref{Equation::EffectiveDensityMatrixElement}.  With this recursive state definition, we can now start the inductive step of the proof.
\subsection{Applying inductive hypothesis}
In order to prove the Identity, we must be able to apply the inductive hypothesis to  Eq. \ref{Equation::BasicSum}.  Ignoring the Clesbsch-Gordan coefficients for the moment, consider an arbitrary term from Eq. \ref{eq:recursiveStateDefinition}.  The dynamics distribute as
	\begin{widetext}
		\begin{align}
				& \sum_{n=1}^N  \sigma_{-}^{(n)} 
				\biggl[ 
				\ketbra{\frac{1}{2},m_i}
				       {\frac{1}{2},m_i'} \otimes \ketbra{J\pm \frac{1}{2},M-m_i,N-1}
				      {J \pm \frac{1}{2},M-m_i',N-1}\biggr]\sigma_{+}^{(n)} \nonumber \displaybreak[1]\\
				= & f^{\frac{1}{2}, m_i,m_i',1}  \otimes \ketbra{J \pm \frac{1}{2},M-m_i,N-1}
				       			{J \pm \frac{1}{2},M-m_i',N-1} + \ketbra{\frac{1}{2},m_i}
				   		 {\frac{1}{2},m_i'}\otimes f^{J  \pm \frac{1}{2},M-m_i,M'-m_i',N-1} \label{eq:recursiveDynamics} \ .
		\end{align}
	\end{widetext}
By extension, all terms in Eq. \ref{eq:recursiveStateDefinition} split the dynamics in this manner, which allows us to apply the inductive hypothesis to evaluate $f^{\frac{1}{2}, m_i,m_i',1}$ and $f^{J \pm \frac{1}{2},M-m_i,M'-m_i',N-1}$.  This means evaluating the $f$ coefficients according to Eq. \ref{Equation::generalFormula}, after which we rewrite the bipartite states in the $N$ spin basis. 
\begin{widetext}
We have the $f^{\frac{1}{2}, m_1,m_1',1}$ terms
\begin{equation}
	\begin{split}
		\frac{1}{d^J_N} \sum_{i_1=1}^{d^{J+\frac{1}{2}}_{N-1}} & \sum_{J_1=J}^{J+1} \sum_{m_1}
			\biggl[ 
			A_q^{\frac{1}{2},m_1}
			\cg{\frac{1}{2}, {m_1}_q}
			   {J + \frac{1}{2}, M-m_1}
			   {J_1, M_q}
			\cg{\frac{1}{2},m_1}
			   {J+\frac{1}{2},M-m_1}
			   {J,M}
			\ket{J_1, M_q; \frac{1}{2}, J + \frac{1}{2}, N-1, i_1}\biggr] \\
			\times &\sum_{J'_1=J}^{J+1}\sum_{m_1'}
			\biggl[ \bra{J'_1, M'_r; \frac{1}{2}, J + \frac{1}{2}, N-1, i_1}
			\cg{\frac{1}{2},m'_1}
			   {J+\frac{1}{2},M'-m'_1}
			   {J,M'}		
			\cg{\frac{1}{2}, {m'_1}_r}
			   {J + \frac{1}{2}, M'-m'_1}
			   {J'_1, M'_r}
			A_r^{\frac{1}{2},m'_1} \biggr]  \label{Equation::Unsimplified::Jplus12::Basic}
   \end{split}
\end{equation}
and the $f^{\frac{1}{2}, m_2,m_2',1}$ terms
\begin{equation}
	\begin{split}
		\frac{1}{d^J_N} \sum_{i_2=1}^{d^{J-\frac{1}{2}}_{N-1}}&\sum_{J_2=J-1}^{J} \sum_{m_2}
				\biggl[ 
				A_q^{\frac{1}{2},m_2}
				\cg{\frac{1}{2}, {m_2}_q}
				   {J - \frac{1}{2}, M-m_2}
				   {J_2, M_q}
				\cg{\frac{1}{2},m_2}
				   {J-\frac{1}{2},M-m_2}
				   {J,M}
				\ket{J_2, M_q; \frac{1}{2}, J - \frac{1}{2}, N-1, i_2}\biggr]\\
		\times & \sum_{J'_2=J-1}^{J}\sum_{m_2'}
		 \biggl[\bra{J'_2, M'_r; \frac{1}{2}, J - \frac{1}{2}, N-1, i_2}
			\cg{\frac{1}{2},m'_2}
			   {J-\frac{1}{2},M'-m'_2}
			   {J,M'}
			\cg{\frac{1}{2}, {m'_2}_r}
			   {J - \frac{1}{2}, M'-m'_2}
			   {J'_2, M'_r}
			A_r^{\frac{1}{2},m_2'}
		\biggr] \ . \label{Equation::Unsimplified::Jminus12::Basic}
	\end{split}
\end{equation}
The $f^{J + \frac{1}{2},M-m_1,M'-m_1',N-1}$ terms are
   \begin{align}
   			\frac{1}{d^J_N(2J+1)}\biggl[1 + 
			\frac{\alpha^{J+\frac{3}{2}}_{N-1}}
			     {d^{J+\frac{1}{2}}_{N-1}}
			\frac{2J+2}
			     {J+\frac{3}{2}}\biggr]
			\sum_{i_1 = 1}^{d^{J+\frac{1}{2}}_{N-1}}
			&\sum_{J_1=J}^{J+1}
			\sum_{m_1}
					\biggl[
					A_q^{J+\frac{1}{2},M-m_1}
					\cg{\frac{1}{2},  m_1}
					   {J + \frac{1}{2}, M_q-m_1}
					   {J_1, M_q}
					\cg{\frac{1}{2},m_1}
					   {J+\frac{1}{2}, M-m_1}
					   {J,M}
					\ket{J_1, M_q; \frac{1}{2}, J+\frac{1}{2}, N-1,i_1}
					\biggr] \nonumber\\
			\times & \sum_{J'_1=J}^{J+1}\sum_{m_1'}
					\biggl[
					\bra{J'_1, M_r'; \frac{1}{2}, J+\frac{1}{2}, N-1,i_1}
					\cg{\frac{1}{2},m'_1}
					   {J+\frac{1}{2}, M'-m'_1}
					   {J,M'}
					\cg{\frac{1}{2},  m'_1}
					   {J + \frac{1}{2}, M'_r-m'_1}
					   {J'_1, M'_r}
					A_r^{J+\frac{1}{2},M'-m'_1}
					\biggr] \displaybreak[1] \label{Equation::Unsimplified::Jplus12::A} \\
	       +\frac{\alpha^{J+\frac{1}{2}}_{N-1}}
				{d^J_N d^{J-\frac{1}{2}}_{N-1}2(J+\frac{1}{2})}
			\sum_{i_1 = 1}^{d^{J-\frac{1}{2}}_{N-1}}
			&\sum_{J_1=J-1}^{J}
			\sum_{m_1}
				\biggl[
				B_q^{J+\frac{1}{2},M-m_1}
				\cg{\frac{1}{2},  m_1}
				   {J - \frac{1}{2}, M_q - m_1}
				   {J_1, M_q}
				\cg{\frac{1}{2},m_1}
				   {J+\frac{1}{2}, M-m_1}
				   {J,M}
				\ket{J_1, M_q; \frac{1}{2}, J-\frac{1}{2}, N-1,i_1}
				\biggr] \nonumber\\
			\times & \sum_{J'_1=J-1}^{J}\sum_{m_1'}
				     \biggl[
					\ket{J'_1, M'_r; \frac{1}{2}, J-\frac{1}{2}, N-1,i_1}					
					\cg{\frac{1}{2},m'_1}
					   {J+\frac{1}{2}, M'-m'_1}
					   {J,M'}
					\cg{\frac{1}{2},  m'_1}
					   {J - \frac{1}{2}, M'_r - m'_1}
					   {J'_1, M'_r}
					B_r^{J+\frac{1}{2},M'-m'_1}
					\biggr] \displaybreak[1] \label{Equation::Unsimplified::Jplus12::B}\\
			+ \frac{\alpha^{J+\frac{3}{2}}_{N-1}}
					{d^J_N d^{J+\frac{3}{2}}_{N-1}2(J+\frac{3}{2})}
		      \sum_{i_1 = 1}^{d^{J+\frac{3}{2}}_{N-1}}
			  &\sum_{J_1=J+1}^{J+2}
			  \sum_{m_1}
				\biggl[
				D_q^{J+\frac{1}{2},M-m_1}
				\cg{\frac{1}{2},  m_1}
				   {J + \frac{3}{2}, M_q - m_1}
				   {J_1, M_q}
				\cg{\frac{1}{2},m_1}
				   {J+\frac{1}{2}, M-m_1}
				   {J,M}
				\ket{J_1, M_q; \frac{1}{2}, J+\frac{3}{2}, N-1,i_1}
				\biggr] \nonumber\\
			\times & \sum_{J'_1=J+1}^{J+2}\sum_{m_1'}
					\biggl[
					\ket{J'_1, M'_r; \frac{1}{2}, J+\frac{3}{2}, N-1,i_1}					
					\cg{\frac{1}{2},m'_1}
					   {J+\frac{1}{2}, M'-m'_1}
					   {J,M'}
					\cg{\frac{1}{2},  m'_1}
					   {J + \frac{3}{2}, M'_r - m'_1}
					   {J'_1, M'_r}
					D_r^{J+\frac{1}{2},M'-m'_1}
					\biggr] \label{Equation::Unsimplified::Jplus12::C}
   		\end{align}
and lastly, the $f^{J-\frac{1}{2},M - m_2, M' - m_2',N-1}$ terms are
	\begin{align}
		\frac{1}{d^J_N2(J-\frac{1}{2})}
		\bigl[ 1 +
		       \frac{\alpha^{J + \frac{1}{2}}_{N-1}}
		            {d^{J-\frac{1}{2}}_{N-1}}
		       \frac{2J}
		            { J + \frac{1}{2}}\bigr]
		\sum_{i_2 = 1}^{d^{J-\frac{1}{2}}_{N-1}}
		&\sum_{J_2=J-1}^{J}
		\sum_{m_2}
			\biggl[
			A_q^{J-\frac{1}{2},M-m_2}
			\cg{\frac{1}{2},  m_2}
			   {J - \frac{1}{2}, M_q - m_2}
			   {J_2, M_q}
			\cg{\frac{1}{2},m_2}
			   {J-\frac{1}{2}, M-m_2}
			   {J,M}
			\ket{J_2, M_q; \frac{1}{2}, J-\frac{1}{2}, N-1,i_2}
			\biggr] \nonumber \\
		\times &\sum_{J_2=J-1}^{J}	\sum_{m'_2}	\biggl[
				\ket{J'_2, M'_r; \frac{1}{2}, J-\frac{1}{2}, N-1,i_2}
				\cg{\frac{1}{2},m'_2}
				   {J-\frac{1}{2}, M'-m'_2}
				   {J,M'}
				\cg{\frac{1}{2},  m'_2}
				   {J - \frac{1}{2}, M'_r - m'_2}
				   {J'_2, M'_r}
				A_r^{J-\frac{1}{2},M'-m'_2}
				\biggr] \displaybreak[1] \label{Equation::Unsimplified::Jminus12::A}\\
		+ \frac{\alpha^{J-\frac{1}{2}}_{N-1}}
					{d^J_N d^{J-\frac{3}{2}}_{N-1}2(J - \frac{1}{2})}
		\sum_{i_2=1}^{d^{J-\frac{3}{2}}_{N-1}}
			&\sum_{J_2=J-2}^{J-1}
		    \sum_{m_2}
			\biggl[
			B_q^{J-\frac{1}{2},M-m_2}
			\cg{\frac{1}{2},  m_2}
			   {J - \frac{3}{2}, M_q - m_2}
			   {J_2, M_q}
			\cg{\frac{1}{2},m_2}
			   {J-\frac{1}{2}, M-m_2}
			   {J,M}
			\ket{J_2, M_q; \frac{1}{2}, J-\frac{3}{2}, N-1,i_2}
			\biggr] \nonumber\\
		\times &\sum_{J_2=J-2}^{J-1}\sum_{m_2'}
			\biggl[
			\ket{J'_2, M'_r; \frac{1}{2}, J-\frac{3}{2}, N-1,i_2}			
			\cg{\frac{1}{2},m'_2}
			   {J-\frac{1}{2}, M'-m'_2}
			   {J,M'}
			\cg{\frac{1}{2},  m'_2}
			   {J - \frac{3}{2}, M'_r - m'_2}
			   {J'_2, M'_r}
			B_r^{J-\frac{1}{2},M'-m'_2}
			\biggr] \displaybreak[1] \label{Equation::Unsimplified::Jminus12::B}\\
		\frac{\alpha^{J+\frac{1}{2}}_{N-1}}
					{d^J_N d^{J+\frac{1}{2}}_{N-1}(2J+1)}
		\sum_{i_2=1}^{d^{J + \frac{1}{2}}_{N-1}}
			&\sum_{J_2=J}^{J+1}
			 \sum_{m_2}
			\biggl[
			D_q^{J-\frac{1}{2},M-m_2}
			\cg{\frac{1}{2},  m_2}
			   {J + \frac{1}{2}, M_q - m_2}
			   {J_2, M_q}
			\cg{\frac{1}{2},m_2}
			   {J-\frac{1}{2}, M-m_2}
			   {J,M}
			\ket{J_2, M_q; \frac{1}{2}, J + \frac{1}{2}, N-1,i_2}
			\biggr] \nonumber\\
		\times & \sum_{J'_2=J}^{J+1}	\sum_{m'_2}
			\biggl[
			\ket{J'_2, M'_r; \frac{1}{2}, J + \frac{1}{2}, N-1,i_2}
			\cg{\frac{1}{2},m'_2}
			   {J-\frac{1}{2}, M'-m'_2}
			   {J,M'}
			\cg{\frac{1}{2},  m'_2}
			   {J + \frac{1}{2}, M'_r - m'_2}
			   {J'_2, M'_r}
			D_r^{J-\frac{1}{2},M-m'_2} 					
			\biggr] \ . \label{Equation::Unsimplified::Jminus12::C}
	\end{align}
\end{widetext}
\subsection{Evaluate sums} \label{SubSubSec::EvaluateSums}

We are now tasked with showing that Eqs. \ref{Equation::Unsimplified::Jplus12::Basic}-\ref{Equation::Unsimplified::Jminus12::C} sum to $f^{J,M,M',N}$ as written in Eq. \ref{Equation::generalFormula}.  Before doing so, we observe that the $J_i,m_i$ and $J'_i,m'_i$ sums factor in all the equations above.  Moreover, if one replaces primed quantities with unprimed ones, the Clebsch-Gordan and $A,B,D$ coefficients of the kets in a given $J_i,m_i$ sum are identical to those of the bras in the related $J'_i,m'_i$ sum.  Therefore, we focus on simplifying the unprimed sums and then apply those results to the primed sums in order to simplify Eqs. \ref{Equation::Unsimplified::Jplus12::Basic}-\ref{Equation::Unsimplified::Jminus12::C}.  In Appendix \ref{Appendix::SimplifySums}, we explicitly calculate two representative sums from these equations.  The calculations involve manipulating products of Clebsch-Gordan and $A,B,D$ coefficients.  Although tedious, the interested and pertinacious reader should have no trouble evaluating them for all relevant sums, finding in particular that the $J\pm 2$ terms vanish.  We forego detailing all those manipulations here and simply use the results in both the primed and unprimed terms of the equations above, which then simplify
\begin{widetext}
\noindent	Eq. \ref{Equation::Unsimplified::Jplus12::Basic} to
	\begin{equation}  \label{Equation::Simplified::JJplus1::First}
		\begin{split}
			\frac{1}{d^J_N(2J+2)^2}\sum_{i_1 = 1}^{d^{J+\frac{1}{2}}_{N-1}}&
			D_q^{J,M}\ketbra{J+1, M_q; \frac{1}{2}, J + \frac{1}{2},N-1, i_1}
			                {J+1, M'_r;\frac{1}{2}, J + \frac{1}{2}, N-1, i_1}D_r^{J,M'}\\
			- & A_q^{J,M}\ketbra{J, M_q; \frac{1}{2}, J + \frac{1}{2}, N-1, i_1}
			                    {J+1, M'_r; \frac{1}{2}, J + \frac{1}{2}, N-1, i_1}D_r^{J,M'}\\
			- & D_q^{J,M}\ketbra{J+1, M_q; \frac{1}{2}, J + \frac{1}{2}, N-1, i_1}
								{J, M'_r; \frac{1}{2}, J + \frac{1}{2}, N-1, i_1}A_r^{J,M'}\\
			+ & A_q^{J,M}\ketbra{J, M_q; \frac{1}{2}, J + \frac{1}{2}, N-1, i_1}
								{J, M'_r; \frac{1}{2}, J + \frac{1}{2}, N-1, i_1}A_r^{J,M'}\ ,
		\end{split}
	\end{equation}	
	Eq. \ref{Equation::Unsimplified::Jminus12::Basic} to 
	\begin{equation} \label{Equation::Simplified::JJminus1::First}
		\begin{split}
			\frac{1}{d^J_N 4J^2}\sum_{i_1=1}^{d^{J-\frac{1}{2}}_{N-1}} &
			  B_q^{J,M} \ketbra{J-1,M_q; \frac{1}{2}, J - \frac{1}{2}; N-1, i_1}
				              {J-1,M_r'; \frac{1}{2}, J - \frac{1}{2}; N-1, i_1} B_r^{J,M'}\\
			 +&A_q^{J,M} \ketbra{J,M_q; \frac{1}{2}, J - \frac{1}{2}; N-1, i_1}
					              {J-1,M_r'; \frac{1}{2}, J - \frac{1}{2}; N-1, i_1} B_r^{J,M'}\\
			 +&B_q^{J,M} \ketbra{J - 1,M_q; \frac{1}{2}, J - \frac{1}{2}; N-1, i_1}
					              {J,M_r'; \frac{1}{2}, J - \frac{1}{2}; N-1, i_1} A_r^{J,M'}\\					
			 +&A_q^{J,M} \ketbra{J,M_q; \frac{1}{2}, J - \frac{1}{2}; N-1, i_1}
					              {J,M_r'; \frac{1}{2}, J - \frac{1}{2}; N-1, i_1} A_r^{J,M'} \ ,
		\end{split}
	\end{equation}
	Eq. \ref{Equation::Unsimplified::Jplus12::A} to 
	\begin{equation}  \label{Equation::Simplified::JJplus1::Second}
		\begin{split}
			\frac{1}{d^J_N(2J+1)} 
			\bigl[1 + 
			\frac{\alpha^{J+\frac{3}{2}}_{N-1}}
			     {d^{J+\frac{1}{2}}_{N-1}} 
			\frac{2J+2}
			     {J+\frac{3}{2}} \bigr]  
			\sum_{i_1=1}^{d^{J+\frac{1}{2}}_{N-1}} 
			&\frac{1}{(2J+2)^2}D_q^{J,M} \ketbra{J+1, M_q;\frac{1}{2}, J + \frac{1}{2}, N-1, i_1}
												{J+1, M'_r;\frac{1}{2}, J + \frac{1}{2}, N-1, i_1}D_r^{J,M'} \\
		   -&\frac{2(J + \frac{3}{2})}{(2J+2)^2} A_q^{J,M} \ketbra{J, M_q;\frac{1}{2}, J + \frac{1}{2}, N-1, i_1}
										{J+1, M'_r;\frac{1}{2}, J + \frac{1}{2}, N-1, i_1}D_r^{J,M'} \\
		   -&\frac{2(J + \frac{3}{2})}{(2J+2)^2} D_q^{J,M}\ketbra{J + 1, M_q;\frac{1}{2}, J + \frac{1}{2}, N-1, i_1}
												{J, M'_r;\frac{1}{2}, J + \frac{1}{2}, N-1, i_1}A_r^{J,M'} \\
		   +&\frac{(J + \frac{3}{2})^2}{(J+1)^2} A_q^{J,M}\ketbra{J, M_q;\frac{1}{2}, J + \frac{1}{2}, N-1, i_1}
													 {J, M'_r;\frac{1}{2}, J + \frac{1}{2}, N-1, i_1}A_r^{J,M'} \ ,
		\end{split}
	\end{equation}
	Eq. \ref{Equation::Unsimplified::Jplus12::B} to
	\begin{equation} \label{Equation::Simplified::JJminus1::Second}
		\begin{split}
			\frac{\alpha^{J+\frac{1}{2}}_{N-1}}
			{d^J_Nd^{J-\frac{1}{2}}_{N-1} 2J(J+\frac{1}{2})}\sum_{i_1=1}^{d^{J-\frac{1}{2}}_{N-1}} &
			   (J+1) B_q^{J,M} \ketbra{J-1,M_q; \frac{1}{2}, J - \frac{1}{2}; N-1, i_1}
				              {J-1,M_r'; \frac{1}{2}, J - \frac{1}{2}; N-1, i_1} B_r^{J,M'}\\
			 +&A_q^{J,M} \ketbra{J,M_q; \frac{1}{2}, J - \frac{1}{2}; N-1, i_1}
					              {J-1,M_r'; \frac{1}{2}, J - \frac{1}{2}; N-1, i_1} B_r^{J,M'}\\
			 +&B_q^{J,M} \ketbra{J - 1,M_q; \frac{1}{2}, J - \frac{1}{2}; N-1, i_1}
					              {J,M_r'; \frac{1}{2}, J - \frac{1}{2}; N-1, i_1} A_r^{J,M'}\\					
			 +&\frac{1}{ J + 1 }A_q^{J,M} \ketbra{J,M_q; \frac{1}{2}, J - \frac{1}{2}; N-1, i_1}
					              {J,M_r'; \frac{1}{2}, J - \frac{1}{2}; N-1, i_1} A_r^{J,M'} \ , 
		\end{split}
	\end{equation}
	Eq. \ref{Equation::Unsimplified::Jplus12::C} to (since $J + 2$ terms vanish) 
	\begin{equation} \label{Equation::Simplified::Jplus1::First}
		\frac{\alpha^{J+1}_N}{d^J_N2(J+1)}
		\frac{1}{d^{J+1}_N}
		\sum_{i_1=1}^{d^{J+\frac{3}{2}}_{N-1}} 
			D_q^{J,M}
			\ketbra{J+1,M_q;\frac{1}{2}, J+\frac{3}{2},N-1,i_1}
			       {J+1,M'_r;\frac{1}{2}, J+\frac{3}{2},N-1,i_1} 
		    D_r^{J,M'} \ ,
	\end{equation}
	Eq. \ref{Equation::Unsimplified::Jminus12::A} to
	\begin{equation} \label{Equation::Simplified::JJminus1::Third}
		\begin{split}
			\frac{1}{d^J_N4J^22(J-\frac{1}{2})}
			\bigl[ 1 +
			       \frac{\alpha^{J + \frac{1}{2}}_{N-1}}
			            {d^{J-\frac{1}{2}}_{N-1}}
			       \frac{2J}
			            { J + \frac{1}{2}}\bigr] &
						\sum_{i_1=1}^{d^{J-\frac{1}{2}}_{N-1}} 
			   B_q^{J,M} \ketbra{J-1,M_q; \frac{1}{2}, J - \frac{1}{2}; N-1, i_1}
				              {J-1,M_r'; \frac{1}{2}, J - \frac{1}{2}; N-1, i_1} B_r^{J,M'}\\
			 -&2(J-\frac{1}{2}) A_q^{J,M} \ketbra{J,M_q; \frac{1}{2}, J - \frac{1}{2}; N-1, i_1}
					              {J-1,M_r'; \frac{1}{2}, J - \frac{1}{2}; N-1, i_1} B_r^{J,M'}\\
			 -&2(J-\frac{1}{2})B_q^{J,M} \ketbra{J - 1,M_q; \frac{1}{2}, J - \frac{1}{2}; N-1, i_1}
					              {J,M_r'; \frac{1}{2}, J - \frac{1}{2}; N-1, i_1} A_r^{J,M'}\\					
			 +&4(J-\frac{1}{2})^2A_q^{J,M} \ketbra{J,M_q; \frac{1}{2}, J - \frac{1}{2}; N-1, i_1}
					              {J,M_r'; \frac{1}{2}, J - \frac{1}{2}; N-1, i_1} A_r^{J,M'}\ ,
		\end{split}
	\end{equation}
	Eq. \ref{Equation::Unsimplified::Jminus12::B} to (since $J - 2$ terms vanish) 
	\begin{equation} \label{Equation::Simplified::Jminus1::First}
		\frac{\alpha^J_N}{d^J_N2J}
		\frac{1}{d^{J-1}_N}
		\sum_{i_1=1}^{d^{J-\frac{3}{2}}_{N-1}}
		B_q^{J,M}
		\ketbra{J-1,M_q;\frac{1}{2},J-\frac{3}{2},N-1,i_1}
		{J-1,M'_r;\frac{1}{2},J-\frac{3}{2},N-1,i_1} 
		B_r^{J,M'} \ ,
	\end{equation}
	and Eq. \ref{Equation::Unsimplified::Jminus12::C} to
	\begin{equation}  \label{Equation::Simplified::JJplus1::Third}
		\begin{split}
			\frac{\alpha^{J+\frac{1}{2}}_{N-1}}
						{d^J_N d^{J+\frac{1}{2}}_{N-1}(2J+1)}
		    \sum_{i_1=1}^{d^{j+\frac{1}{2}}_{N-1}} 
			&\frac{J}{J+1}D_q^{J,M}\ketbra{J+1, M_q;\frac{1}{2}, J + \frac{1}{2}, N-1, i_1}
					{J+1, M'_r;\frac{1}{2}, J + \frac{1}{2}, N-1, i_1}D_r^{J,M'} \\
		   +&\frac{1}{J+1}A_q^{J,M}\ketbra{J, M_q;\frac{1}{2}, J + \frac{1}{2}, N-1, i_1}
						{J+1, M'_r;\frac{1}{2}, J + \frac{1}{2}, N-1, i_1}D_r^{J,M'} \\			
		   +&\frac{1}{J+1}D_q^{J,M}\ketbra{J+1, M_q;\frac{1}{2}, J + \frac{1}{2}, N-1, i_1}
									{J, M'_r;\frac{1}{2}, J + \frac{1}{2}, N-1, i_1}A_r^{J,M'} \\					
		   +&\frac{1}{J(J+1)}A_q^{J,M}\ketbra{J, M_q;\frac{1}{2}, J + \frac{1}{2}, N-1, i_1}
						{J, M'_r;\frac{1}{2}, J + \frac{1}{2}, N-1, i_1}A_r^{J,M'} \ .
		\end{split}
	\end{equation}
\end{widetext} 
\subsection{Recover $f^{J,M,M',N}$}
We now combine the equations from the previous subsection to recover the Identity in Eq. \ref{Equation::generalFormula}.  Given that density operators in the collective state representation lack coherences between different $J$ irreps, we expect $\ketbra{J\pm 1}{J}$ and $\ketbra{J}{J \pm 1}$ terms to vanish.  Since both the $\ketbra{J}{J \pm 1}$ and $\ketbra{J \pm 1}{J}$ terms have the same coefficients, we need only explicitly deal with one of the two.  Starting with $\ketbra{J+1}{J}$ coefficients from Eqs. \ref{Equation::Simplified::JJplus1::First}, \ref{Equation::Simplified::JJplus1::Second} and \ref{Equation::Simplified::JJplus1::Third}, we find
		\begin{align} 
				\frac{1}{d^J_N}&\biggl(
				-\frac{1}{(2J+2)^2} -\frac{(2J+3)}{(2J+1)(2J+2)^2}\bigl[1 + 
				           \frac{\alpha^{J+\frac{3}{2}}_{N-1}}
				                {d^{J+\frac{1}{2}}_{N-1}}
				    \frac{2J+2}{J+\frac{3}{2}}\bigr] \nonumber\\
				&+\frac{1}{(J+1)(2J+1)}
				 \frac{\alpha^{J+\frac{1}{2}}_{N-1}}
				      {d^{J+\frac{1}{2}}_{N-1}}\biggr)\nonumber\displaybreak[1]\\
				= &\frac{1}{d^J_N(2J+2)^2}\biggl(-1-\frac{(N+1)}{2J+1}+\frac{N+2J+2}{2J+1}\biggr) \\
			    = &\ 0
		\end{align}
Similarly, for $\ketbra{J-1}{J}$ coefficients in Eqs. \ref{Equation::Simplified::JJminus1::First}, \ref{Equation::Simplified::JJminus1::Second} and \ref{Equation::Simplified::JJminus1::Third}, we have
\begin{equation}
   \frac{1}{d^J_N4J^2}\biggl[ 1 + 
					\frac{\alpha^{J+\frac{1}{2}}_{N-1} 2J}
					{d^{J-\frac{1}{2}}_{N-1}(J+\frac{1}{2})}
					- \bigl[ 1 +
					       \frac{\alpha^{J + \frac{1}{2}}_{N-1}}
					            {d^{J-\frac{1}{2}}_{N-1}}
					       \frac{2J}
					            { J + \frac{1}{2}}\bigr]\biggr] = 0
\end{equation}

Turning to $J+1$ terms from Eqs. \ref{Equation::Simplified::JJplus1::First}, \ref{Equation::Simplified::JJplus1::Second} and \ref{Equation::Simplified::JJplus1::Third}, the coefficients sum to
\begin{align}
		\frac{1}{d^J_N}  &
		\biggl(\frac{1}{(2J+2)^2} + 
		      \frac{1}{(2J+1)(2J+2)^2} 
		         \bigl[1 + 
		           \frac{\alpha^{J+\frac{3}{2}}_{N-1}}
		                {d^{J+\frac{1}{2}}_{N-1}}
		           \frac{2J+2}{J+\frac{3}{2}}\bigr] \\
		       & + \frac{J}{(J+1)(2J+1)}
		         \frac{\alpha^{J+\frac{1}{2}}_{N-1}}
		          {d^{J+\frac{1}{2}}_{N-1}}\biggr)\nonumber\displaybreak[1]\\
		=&     \frac{1}{d^J_N(2J+2)^2} 
		     \biggl(1+\frac{N+1}{(2J+3)(2J+1)} 
		         + \frac{J(N+2J+2)}{2J+1}\biggr)\nonumber\displaybreak[1]\\
		=&\frac{1}{d^J_N} 
		 \frac{2 J+N+4}{8 J^2+20 J+12}\nonumber\\
		=& \frac{1}{d^J_N2(J+1)}
		    \frac{\alpha^{J+1}_N}{d^{J+1}_N}
\end{align}
which gives overall
	\begin{multline}  \label{Equation::Simplified::Jplus1::Second}
			\frac{\alpha^{J+1}_N}{d^J_N2(J+1)}
			\frac{1}{d^{J+1}_N}
			\sum_{i_1=1}^{d^{J+\frac{1}{2}}_N}
			D_q^{J,M}
			\ket{J+1,M_q;\frac{1}{2},J+\frac{1}{2},N-1,i_1}\\
			       \times\bra{J+1,M'_r;\frac{1}{2},J+\frac{1}{2},N-1,i_1}
			D_r^{J,M'} \ .
	\end{multline}
	
The $J$ terms from Eqs. \ref{Equation::Simplified::JJplus1::First}, \ref{Equation::Simplified::JJplus1::Second} and \ref{Equation::Simplified::JJplus1::Third} have coefficients
	\begin{align}
			\frac{1}{d^J_N} &
			\biggl(\frac{1}{(2J+2)^2} 
			 + \frac{(2J+3)^2}{(2J+1)(2J+2)^2}
			    \bigl[1 + 
			          \frac{\alpha^{J+\frac{3}{2}}_{N-1}}
			               {d^{J+\frac{1}{2}}_{N-1}}
			          \frac{2J+2}{J+\frac{3}{2}}\bigr] \nonumber\\
			 & + \frac{1}{J(J+1)(2J+1)}
			    \frac{\alpha^{J+\frac{1}{2}}_{N-1}}
			      {d^{J+\frac{1}{2}}_{N-1}}\biggr) \nonumber \displaybreak[1]\\
			=&\frac{1}{d^J_N(2J+2)^2}
			    \biggl(1+\frac{(N+1)(2J+3)}{2J+1}
			   +\frac{N+2J+2}{J(2J+1)}\biggr) \\ 
		   =& \frac{1}{d^J_N2J} \biggl[ 1 + 
			    \frac{\alpha^{J+1}_N}{d^J_N}
			     \frac{2J+1}{J+1}\biggr]
	\end{align}
which gives overall
	\begin{multline} \label{Equation::Simplified::J::First}
	  \frac{1}{2J}
	  \biggl[1+
	         \frac{\alpha^{J+1}_N}{d^J_N}
	         \frac{2J+1}{J+1}\biggr] \times\\
	         \frac{1}{d^J_N}\sum_{i_2=1}^{d^{J+\frac{1}{2}}_{N-1}}
			 A_q^{J,M}
	         \ket{J,M_q;\frac{1}{2},J+\frac{1}{2},N-1,i_1}\\
	            \times  \bra{J,M'_r;\frac{1}{2},J+\frac{1}{2},N-1,i_1}
	         A_r^{J,M'} \ .
	\end{multline}

Similarly, the $J$ terms from Eqs. \ref{Equation::Simplified::JJminus1::First}, \ref{Equation::Simplified::JJminus1::Second} and \ref{Equation::Simplified::JJminus1::Third} have coefficients
\begin{align}
		\frac{1}{d^J_N4J^2}&\biggl[ 1 + 
							\frac{\alpha^{J+\frac{1}{2}}_{N-1} 2J}
							{d^{J-\frac{1}{2}}_{N-1}(J+\frac{1}{2})(J+1)} \nonumber\\
							&+ 2(J-\frac{1}{2})\bigl[ 1 +
							       \frac{\alpha^{J + \frac{1}{2}}_{N-1}}
							            {d^{J-\frac{1}{2}}_{N-1}}
							       \frac{2J}
							            { J + \frac{1}{2}}\bigr]\biggr] \nonumber\displaybreak[1]\\
		= & \frac{1}{d^J_N4J^2}\biggl[ 2J + \frac{N - 2J}{2J+1}\bigl(\frac{1}{J+1} + 2J -1\bigr)\biggr]\nonumber\\
		=& \frac{1}{d^J_N2J}\biggl[ 1 + \frac{\alpha^{J+1}_{N}}
											 {d^J_{N}}
										\frac{2J+1}
										     {J+1}\biggr]
\end{align}
which gives 
\begin{multline} \label{Equation::Simplified::J::Second}
	\frac{1}{2J}\biggl[ 1 + \frac{\alpha^{J+1}_{N}}
										 {d^J_{N}}
									\frac{2J+1}
									     {J+1}\biggr] \times\\
	\frac{1}{d^J_N}\sum_{i_1 = 1}^{d^{J - \frac{1}{2}}_{N-1}}
	A_q^{J,M}
	\ket{J,M_q;\frac{1}{2},J-\frac{1}{2},N-1,i_1}\\
	\times	 \bra{J,M'_r;\frac{1}{2},J-\frac{1}{2},N-1,i_1}
	A_r^{J,M'} \ .
\end{multline}

And finally, the $J-1$ sums from Eqs. \ref{Equation::Simplified::JJminus1::First}, \ref{Equation::Simplified::JJminus1::Second} and \ref{Equation::Simplified::JJminus1::Third} have coefficients
	\begin{align} 
			\frac{1}{d^J_N4J^2}&\biggl[ 1 + 
								\frac{\alpha^{J+\frac{1}{2}}_{N-1} 2J(J+1)}
								{d^{J-\frac{1}{2}}_{N-1}(J+\frac{1}{2})} \nonumber\\
								&+ \frac{1}{2(J-\frac{1}{2})}\bigl[ 1 +
								       \frac{\alpha^{J + \frac{1}{2}}_{N-1}}
								            {d^{J-\frac{1}{2}}_{N-1}}
								       \frac{2J}
								            { J + \frac{1}{2}}\bigr]\biggr] \nonumber\displaybreak[1]\\
		  &=\frac{1}{d^J_N4J^2}\biggl[ 1 + \frac{1}{2J-1}  + 
		                               \frac{N-2J}{2J+1}\bigl( J + 1 + \frac{1}{2J-1}\bigr) \biggr] \nonumber\\
		  &=\frac{1}{d^J_N2J}\frac{\alpha^J_N}{d^{J-1}_N}		
	\end{align}
which gives
	\begin{multline} \label{Equation::Simplified::Jminus1::Second}
		\frac{\alpha^J_N}{d^J_N2J}\frac{1}{d^{J-1}_N}
			\sum_{i_1=1}^{d^{J-\frac{1}{2}}_{N-1}}
			B_q^{J,M}
			\ket{J - 1,M_q;\frac{1}{2},J-\frac{1}{2},N-1,i_1}\\
			\times	   \bra{J - 1,M'_r;\frac{1}{2},J-\frac{1}{2},N-1,i_1}
			B_r^{J,M'} \ .
	\end{multline}
	
From the definition of $\ketbra{J,M,N}{J,M',N}$ given in Eq. \ref{Equation::EffectiveDensityMatrixElement}, we see that Eqs. \ref{Equation::Simplified::J::First} and \ref{Equation::Simplified::J::Second} correspond to the $\ketbra{J,M,N}{J,M',N}$ terms in Eq. \ref{Equation::generalFormula}.  A similar combination of Eqs. \ref{Equation::Simplified::Jminus1::First} and \ref{Equation::Simplified::Jminus1::Second} corresponds to the $J-1$ term and the combination of Eqs. \ref{Equation::Simplified::Jplus1::First} and \ref{Equation::Simplified::Jplus1::Second} corresponds to the $J$ term.  We have thus shown inductively that Identity \ref{Id::MainResult} holds. $\blacksquare$
\section{Conclusion} \label{Section::Conclusion}
We have presented an exact formula for efficiently expressing symmetric processes of an ensemble of spin-1/2 particles.  The efficiency is achieved by generalizing the notion of collective spin states to be any such state which does not distinguish degenerate irreps.  For a collection of $N$ spin-1/2 particles, the effective Hilbert space dimension grows as $N^2$, a drastic reduction from the full Hilbert space scaling of $2^N$.  The collective representation is used in Identity \ref{Id::MainResult}, which gives a closed-form expression for evaluating non-collective terms from symmetric Lindblad operators.  Unfortunately, due to the complicated structure of adding spin-$J>\frac{1}{2}$ particles \cite{Mihailov1977}, our results do not appear to generalize.  Nonetheless, we believe that this approach will become a useful tool in analyzing collective spin phenomenon and in particular, accurately considering the role of decoherence in collective spin experiments \cite{shortPaper}.

\appendix
\section{Explicit Simplification of Typical Sums} \label{Appendix::SimplifySums}

	In \ref{SubSubSec::EvaluateSums}, we simplify the sums in Eqs. \ref{Equation::Unsimplified::Jplus12::Basic}-\ref{Equation::Unsimplified::Jminus12::C} but do not go through the detailed algebra.  The work involves manipulating products of Clebsch-Gordan and $A,B,D$ coefficients.  In this appendix, we explicitly calculate two representative sums from this set and invite the reader to calculate the remainder in a similar fashion.  
	
	First, consider the sums over $J_1$ and $m_1$ in Eq. \ref{Equation::Unsimplified::Jplus12::Basic}, which is representative of sums in  Eqs. \ref{Equation::Unsimplified::Jplus12::Basic} and \ref{Equation::Unsimplified::Jminus12::Basic}.  For $J_1 = J+1$
	\begin{align}
		&A_q^{\frac{1}{2},\frac{1}{2}}
		\cg{\frac{1}{2}, {\frac{1}{2}}_q}
		   {J + \frac{1}{2}, M-\frac{1}{2}}
		   {J+1, M_q}
		\cg{\frac{1}{2},\frac{1}{2}}
		   {J+\frac{1}{2},M-\frac{1}{2}}
		   {J,M}\nonumber\\
	    &+
		A_q^{\frac{1}{2},-\frac{1}{2}}
		\cg{\frac{1}{2}, {-\frac{1}{2}}_q}
		   {J + \frac{1}{2}, M + \frac{1}{2}}
		   {J+1, M_q}
		\cg{\frac{1}{2},-\frac{1}{2}}
		   {J+\frac{1}{2},M + \frac{1}{2}}
		   {J,M}\nonumber \displaybreak[1]\\
		=&\frac{1}{2(J+1)} \begin{cases}
			    -\sqrt{(J+M+2)(J+M+1)} & q = +\\
			     \sqrt{(J-M+2)(J-M+1)} & q = -\\
			     \sqrt{(J-M+1)(J+M+1)} & q = z
	   	    \end{cases}\\
		=&  \frac{1}{2J+2} D_q^{J,M}
	\end{align}
and for $J_1 = J$
	\begin{align}
		&A_q^{\frac{1}{2},\frac{1}{2}}
		\cg{\frac{1}{2}, {\frac{1}{2}}_q}
		   {J + \frac{1}{2}, M-\frac{1}{2}}
		   {J, M_q}
		\cg{\frac{1}{2},\frac{1}{2}}
		   {J+\frac{1}{2},M-\frac{1}{2}}
		   {J,M}\nonumber\\
	    &+ 
		A_q^{\frac{1}{2},-\frac{1}{2}}
		\cg{\frac{1}{2}, {-\frac{1}{2}}_q}
		   {J + \frac{1}{2}, M + \frac{1}{2}}
		   {J, M_q}
		\cg{\frac{1}{2},-\frac{1}{2}}
		   {J+\frac{1}{2},M + \frac{1}{2}}
		   {J,M}\nonumber \displaybreak[1]\\
		=&-\frac{1}{2(J+1)}
			\begin{cases}
				\sqrt{(J-M)(J+M+1)} & q = +\\
				\sqrt{(J+M)(J-M+1)} & q = -\\
				M & q = z
			\end{cases} \\
		= & -\frac{1}{2J+2} A_q^{J,M}
	\end{align}
	where $A_{+}^{\frac{1}{2},\frac{1}{2}} = A_{-}^{\frac{1}{2},-\frac{1}{2}} = 0$, $A_{+}^{\frac{1}{2},-\frac{1}{2}} = A_{-}^{\frac{1}{2},\frac{1}{2}} = 1$ and $A_{z}^{\frac{1}{2},\pm \frac{1}{2}} = \pm \frac{1}{2}$. 
	
Similarly, consider the sums over $J_1$ and $m_1$ in Eq. \ref{Equation::Unsimplified::Jplus12::B}, which is representative of Eqs. \ref{Equation::Unsimplified::Jplus12::A}-\ref{Equation::Unsimplified::Jminus12::C}.  For $J_1 = J-1$, we have
	\begin{align}
		B_q^{J+\frac{1}{2},M-\frac{1}{2}}&
		\cg{\frac{1}{2},  \frac{1}{2}}
		   {J - \frac{1}{2}, M_q - \frac{1}{2}}
		   {J-1, M_q}
		\cg{\frac{1}{2},\frac{1}{2}}
		   {J+\frac{1}{2}, M-\frac{1}{2}}
		   {J,M} \nonumber\\
	   &+B_q^{J+\frac{1}{2},M + \frac{1}{2}}
		\cg{\frac{1}{2},  -\frac{1}{2}}
		   {J - \frac{1}{2}, M_q + \frac{1}{2}}
		   {J-1, M_q}
		\cg{\frac{1}{2},-\frac{1}{2}}
		   {J+\frac{1}{2}, M + \frac{1}{2}}
		   {J,M} \nonumber \displaybreak[1]\\
	 = & \sqrt{\frac{(J-M_q)(J-M+1)}
			        {4J(J+1)}}
			B_q^{J+\frac{1}{2},M-\frac{1}{2}}\nonumber\\
		&\times \biggl[ 1 + 	\sqrt{\frac{J+M_q}
					   {J-M_q}}
			\sqrt{\frac{J+M+1}
					   {J-M+1}}
			\frac{B_q^{J+\frac{1}{2},M + \frac{1}{2}}}
			     {B_q^{J+\frac{1}{2},M-\frac{1}{2}}}\biggr]\nonumber\displaybreak[1]\\
	 = & \sqrt{\frac{(J-M_q)(J-M+1)}
			        {4J(J+1)}}
			B_q^{J+\frac{1}{2},M-\frac{1}{2}}\frac{2(J+1)}{J-M+1}\nonumber\displaybreak[1]\\
	 = & \sqrt{\frac{J+1}{J}}
		\begin{cases}
			\sqrt{(J-M)(J-M-1)} & q = +\\
		   -\sqrt{(J+M)(J+M-1)} & q = -\\
		    \sqrt{(J+M)(J-M)} &   q = z
	 	 \end{cases} \nonumber\\
	 = & \sqrt{\frac{J+1}{J}} B_q^{J,M} \ .
	\end{align}
Similarly, for $J_1 = J$, we have
	\begin{align}
		B_q^{J+\frac{1}{2},M-\frac{1}{2}}&
		\cg{\frac{1}{2},  \frac{1}{2}}
		   {J - \frac{1}{2}, M_q - \frac{1}{2}}
		   {J, M_q}
		\cg{\frac{1}{2},\frac{1}{2}}
		   {J+\frac{1}{2}, M-\frac{1}{2}}
		   {J,M} \nonumber \\
	   &+ B_q{J+\frac{1}{2},M + \frac{1}{2}}
		\cg{\frac{1}{2},  -\frac{1}{2}}
		   {J - \frac{1}{2}, M_q + \frac{1}{2}}
		   {J, M_q}
		\cg{\frac{1}{2},-\frac{1}{2}}
		   {J+\frac{1}{2}, M + \frac{1}{2}}
		   {J,M} \nonumber\displaybreak[1]\\
		= & \sqrt{\frac{(J + M_q)(J-M+1)}
				        {4J(J+1)}}
				B_q^{J+\frac{1}{2},M-\frac{1}{2}} \nonumber\\
		  &\times		\biggl[ 1 -
				\sqrt{\frac{J-M_q}
						   {J+M_q}}
				\sqrt{\frac{J+M+1}
						   {J-M+1}}
				\frac{B_q^{J+\frac{1}{2},M + \frac{1}{2}}}
				     {B_q^{J+\frac{1}{2},M-\frac{1}{2}}}\biggr]\nonumber\displaybreak[1]\\
	    = & \sqrt{\frac{(J + M_q)(J-M+1)}
				        {4J(J+1)}}
				B_q^{J+\frac{1}{2},M-\frac{1}{2}} \nonumber\\
		  & \times 
			2\begin{cases}
				\frac{1}{J-M+1} & q = +\\
			   -\frac{1}{J + M + 1} & q = -\\
			    \frac{M}{(J+M)(J-M+1)} & q = z
			\end{cases} \nonumber \displaybreak[1]\\
		= & \sqrt{\frac{1}{J(J+1)}}
			\begin{cases}
				\sqrt{(J-M)(J+M+1)} & q = +\\
				\sqrt{(J+M)(J-M+1)} & q = -\\				
				M & q = z				
			\end{cases} \nonumber \displaybreak[1]\\
		= & \sqrt{\frac{1}{J(J+1)}} A_q^{J,M} \ .
	\end{align}
\begin{acknowledgments}
This work was inspired by discussions with Ivan Deutsch and Poul Jessen and was supported by the NSF (PHY-0652877) and the DOE under contract with Sandia National Laboratory and the NINE program (707291).
\end{acknowledgments}

\end{document}